\documentclass[12pt]{iopart}

\usepackage{graphicx}
\usepackage{subfigure}
\begin{document}

\title{Non-linear propagation effects of intense femtosecond pulses on low order harmonics in solids}

\author{M Hussain $^{1}$, G O Williams $^{1}$, T Imran $^{2}$ and M  Fajardo $^{1}$ }

\address{$^{1}$ GoLP/Instituto de Plasmas e Fusao Nuclear-Laboratorio Associado, Instituto Superior Tecnico, Universidade de Lisboa, 1049-001 Lisboa, Portugal\\
	$^{2}$ Research Laboratory of Lasers (RLL)-Group of Laser Development (GoLD), Department of Physics, COMSATS University Islamabad, Park Road, 44000, Islamabad, Pakistan}
\ead{mukhtar.hussain@tecnico.ulisboa.pt}
\vspace{10pt}

\begin{abstract}
The non-linear propagation of the intense near-infrared (NIR) driving field in wide bandgap materials pose a challenge and an opportunity to control the spectral properties of high harmonic generation (HHG) in solids. Here, we have investigated the non-linear propagation effects of the ultrafast intense near-infrared (NIR) driving field at 800 nm of 40 fs pulse duration operating at a repetition rate of 1 kHz focused on the wide bandgap dielectrics such as MgO, Chromium (Cr) doped MgO (Cr: MgO), Sapphire (Sa) crystals and fused silica (FS). Furthermore, we have generated second and third harmonic (TH) in these materials to explore the non-linear response at a strong field. To quantify the non-linear propagation effects, low-order harmonics have been generated in reflection and compared with the harmonics generated in transmission. 
We observe spectral shifts and broadening of the driving field spectrum which is imprinted on the harmonics. We attribute these effects to strong photoionization, generation of free-carrier density and self-phase modulation effects. We have also studied the polarization dependence of second harmonic generation (SHG) and TH in FS. The linear polarization dependence of below bandgap harmonics in FS and Sa generated in reflection demonstrated the sharp anisotropy than in transmission. This work shows the sensitivity to control the spectral profile of harmonics by manipulating the driving field, showing the possibility of new tailored solid-state XUV sources for optical diagnostics.
\end{abstract}
%
%
%
%
%

\section{Introduction}
The microscopic origin of high harmonic generation (HHG) in wide bandgap crystalline and amorphous solids have been reported and offering a promising route for new all-solid-state extreme-ultraviolet sources \cite{you2017anisotropic,Han,you2017high}. The non-linear optical response of solids can be manipulated by either tailoring the electronic structures \cite{dmitriev2013handbook,hussain2020controlling} or by the propagation of intense driving field in thin solids  \cite{hussain2021spectral} which propose the propagation effects an important factor when generating harmonics in solids. The propagation effects of the intense femtosecond pulses in gases and plasma induce the red-shifted harmonics \cite{brandi2006spectral,bian2013spectral}. Recently, we have demonstrated non-linear macroscopic beam propagation effects on harmonics in semiconductors such as  Si and ZnO. We have observed the spectral shifts in the generated harmonics which is attributed to the strong photoionization of the valence band through non-linear beam propagation \cite{hussain2021spectral}. 

The spatial mode characterization of the fundamental beam through pristine MgO and Cr: MgO showed the prominent self-focusing effect in crystals \cite{Shathathesis}. The driving pulse-modulated through the non-linear propagation effects, which eventually generate harmonics during propagation in the medium. As a consequence of non-linear propagation, these effects can be imprinted on the harmonics.  This study is limited to the spatial mode characterization at a longer wavelength (1.78 $\mu$m) and showed harmonics spatial mode contraction in Cr: MgO compared to pristine MgO\cite{Shathathesis}. Similarly, the extreme ultraviolet (EUV) generated in Sa and steered the propagation of EUV by simply shaping the crystal surface \cite{kim2017generation}. Yet, the propagation of the driving beam centred at 800 nm in high bandgap solids and its impact on the below bandgap harmonics is often overlooked.

Here, we have measured the propagation effects of the driving field in MgO, Cr: MgO, Sa and FS at 800 nm, 40 fs operating at 1 kHz by measuring the spectral profile. The influence of doping concentration of Cr on the propagated fundamental beam in Cr: MgO is investigated.  We have observed the blue-shifted broadening of the fundamental spectrum in MgO due to strong photoionization and overall broadening in higher doped MgO crystals attributed to the self-phase modulation. Similarly, we have observed the broadening of the fundamental spectrum through FS both in transmission and reflection while in Sa, the FWHM of the fundamental spectrum reduced by 3.3 nm in transmission. In addition, the effect of non-linear propagation on below bandgap harmonics (2nd and 3rd) have been demonstrated. The polarization response of below bandgap harmonics in transmission and reflection geometry in Sa, SiO$_2$ and FS showed more anisotropy in reflection than in transmission geometry which we attribute to non-linear propagation effect.

This article is structured as, in section \ref{exp:setup_PST} the experimental setup to measure propagation effects and to generate below bandgap harmonics is presented. The propagation effects of driving pulses and third harmonic generation (THG) in MgO and Cr: MgO is illustrated in section \ref{IRharmonics: MgO and CRMgO} and \ref{THG:MgO and CrMgO}, respectively. In section \ref{SHG_THG:Sa} and \ref{SHG_THG:FS}, propagation effects of driving pulses and generation and polarization dependence of harmonics in Sa and FS is discussed. Finally, the conclusion of this work is reported in section \ref{conclusion}.

\section{Experimental setup}
\label{exp:setup_PST}

\begin{figure}[htbp]
   		\subfigure[\label{fig:Schemtic_PST}]{{\includegraphics[scale=0.25]{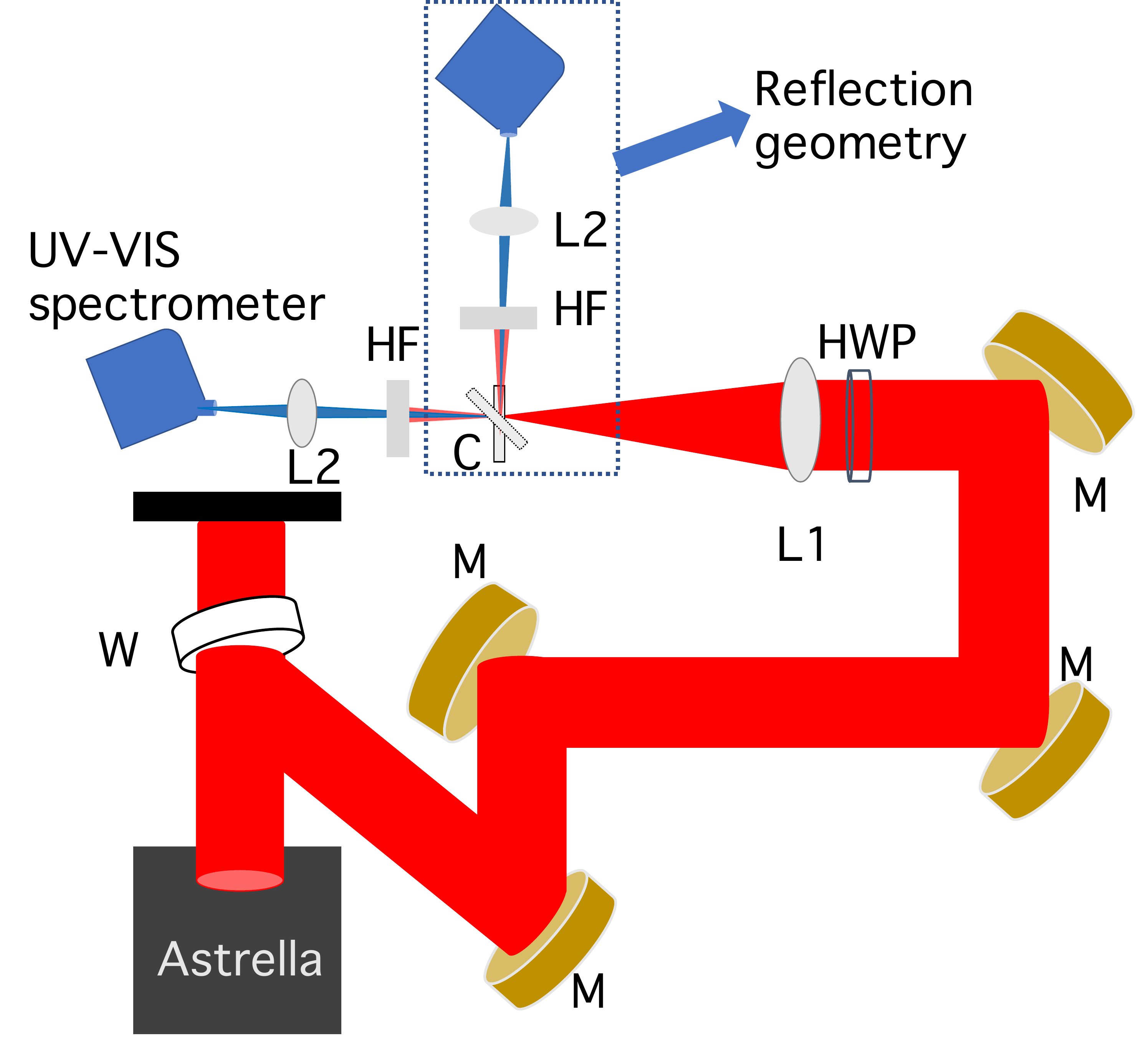}}}
		\subfigure[\label{fig:IR_trans_Cr_MgO}]{{\includegraphics[scale=0.25]{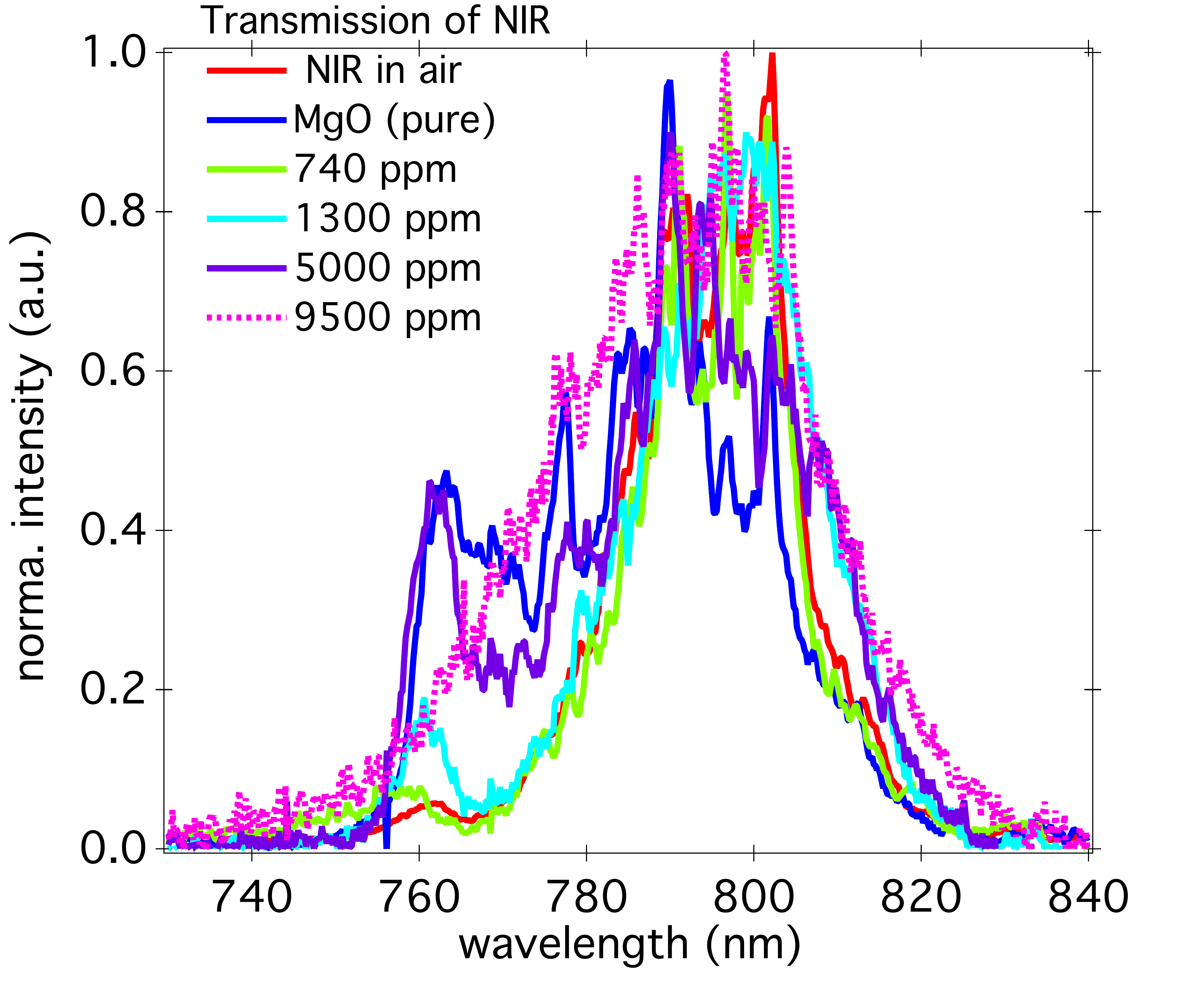}}}
	\caption{(a) Schematic of the experimental setup to observe the propagation effects and to generate below bandgap harmonics in transmission and reflection geometry in thin solids. For reflection geometry, solids are aligned to 45$^\circ$ to the incoming driving laser pulses and reflected the signals which are filtered by HF and focused to UV-VIS spectrometer by L2. (b) Propagation of NIR (800 nm) in 200 $\mu$m  MgO and Cr: MgO crystals. Where HWP: half-wave plate,  W: wedge, L1: 75 cm focal length, L2: 10 cm focal length, C: crystal, HF: harmonic filter, M: NIR-reflecting mirror.}
	
\end{figure}

The schematic of the experimental setup to observe SHG and THG in thin solids in transmission and reflection geometry is shown in figure \ref{fig:Schemtic_PST} . We have used the Astrella femtosecond laser and insert a wedge (W) to reflect an average power of 20 mW to execute the experiment. The NIR driving pulses of central wavelength 800 nm, 40 fs \cite{hussain2020controlling}, operating at 1kHz are focused on the crystals to observe the propagation effects in the MgO (pure) and Cr: MgO, Sa and FS in transmission geometry. The NIR pulses are focused to thin solids by a lens (L1) of 75 cm focal length  to $\sim$ 100 $\mu$m spot size having an estimated peak intensity $\sim$ 1.0 $\times$ $10^{13}$ W/cm\textsuperscript{2}. The propagated and reflected NIR pulses through solids are measured using a UV-VIS spectrometer. Furthermore, below band-gap harmonics (SHG and THG) have been generated at the estimated peak intensity $\sim$ 1.0 $\times$ $10^{13}$ W/cm\textsuperscript{2} in solids in transmission and reflection geometry. The generated harmonics have been filtered with the corresponding harmonic filters (HF). The filtered signal of harmonics are focused to the UV-VIS spectrometer by a lens (L2) of 10 cm focal length and recorded in transmission (T) and reflection (R) geometry (figure \ref{fig:Schemtic_PST} ). Furthermore, we have employed a half-wave plate (HWP) to measure the polarization response of below bandgap harmonics in transmission and reflection geometry in Sa, SiO$_2$ and FS.

\section{Result and discussions}
\subsection{Propagation effect of NIR in MgO and Cr: MgO}
\label{IRharmonics: MgO and CRMgO}

The spectrum of driving field NIR centered at 800 nm and its transmission through the MgO (pure) and Cr: MgO crystals at an intensity of $\sim$1.0 $\times$ $10^{13}$ W/cm\textsuperscript{2} is shown in figure \ref{fig:IR_trans_Cr_MgO}. The Gaussian fit FWHM of NIR in the absence of crystals is 23.8 nm while FWHM of NIR through MgO is 40.96 nm as shown in figure \ref{fig:IR_trans_CrMgO}. We have observed the broadening and blue-shifting in the transmitted spectra as seen in the Gaussian fit (red dotted curve of NIR and the blue dotted curve represents the transmission of NIR through MgO). With the increase of the doping concentration of Cr in MgO, the transmitted spectra of NIR broaden as shown in figures \ref{fig:IR_trans_CrMgO}. The FWHM of NIR transmitted through Cr: MgO (5000 ppm and 9500 ppm) increased to 35.3 nm and 38.5 nm, respectively.

The amount of blue-shifted spectral broadening in MgO (figure \ref{fig:IR_trans_Cr_MgO}) is limited due to the multiphoton absorption process, counterbalance of free-carrier density and Kerr nonlinearity at the leading edge and attenuation experienced by the components of blue-shifted frequency at the trailing edge of the optical pulse \cite{koonath2008limiting}. As the refractive index of the medium modulated due to the Kerr effect and free-carrier density due to the self-focusing effect and multiphoton absorption process. The Kerr non-linearity induces the blue-shift in the spectrum of a pulse at the trailing edge and redshift at the leading edge while the increase in free-carrier density induces blue shift at both edges of the pulse \cite{koonath2008limiting}. Thus, we observe a net blue-shifted spectrum in MgO. The self-focusing effect is observed in Cr: MgO crystals \cite{Shathathesis} which induces the self-phase modulation (SPM). SPM is a non-linear effect that induces the varying refractive index with the interaction of light with the medium. The variation in the refractive index of the medium produces the phase shift and as result, the spectrum of the pulse changed.  We have observed broaden transmitted driving spectra through Cr: MgO (figure \ref{fig:IR_trans_Cr_MgO}). Our results show that the transmission of the driving field decreases with the increase of doping concentration, particularly for 9500 ppm dopant concentration. Therefore, the conversion efficiency of NIR photon to UV photon will be smaller in Cr: MgO crystals compared to pure MgO.

\subsection{Generation and transmission of TH in MgO and Cr: MgO crystals }
\label{THG:MgO and CrMgO}

\begin{figure}[htbp]
	\subfigure[\label{fig:THG_SpectrumT_CrMgO}]{{\includegraphics[scale=0.19]{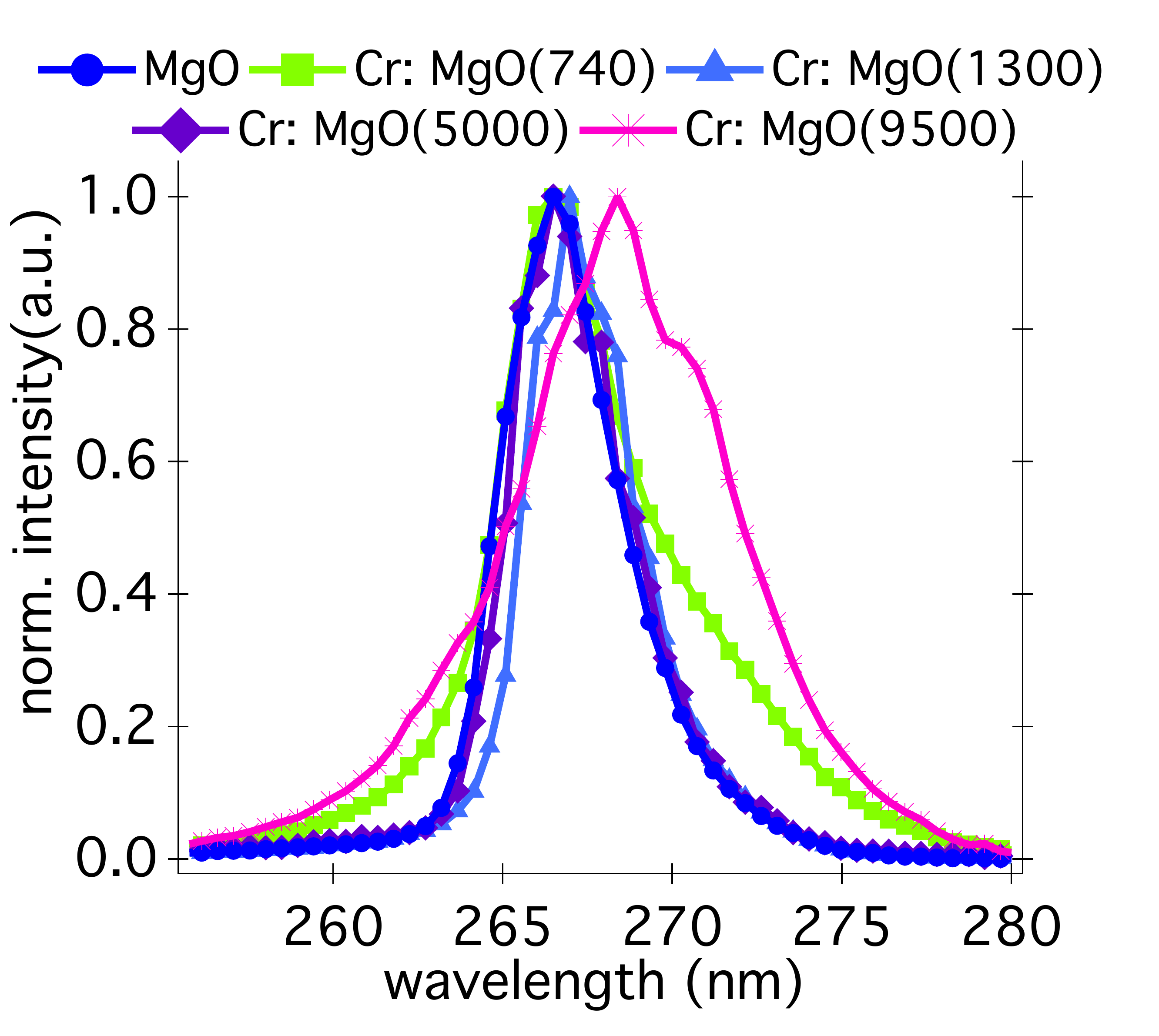}}}
	\subfigure[\label{fig:THG_trans_Ref_MgO}]{{\includegraphics[scale=0.19]{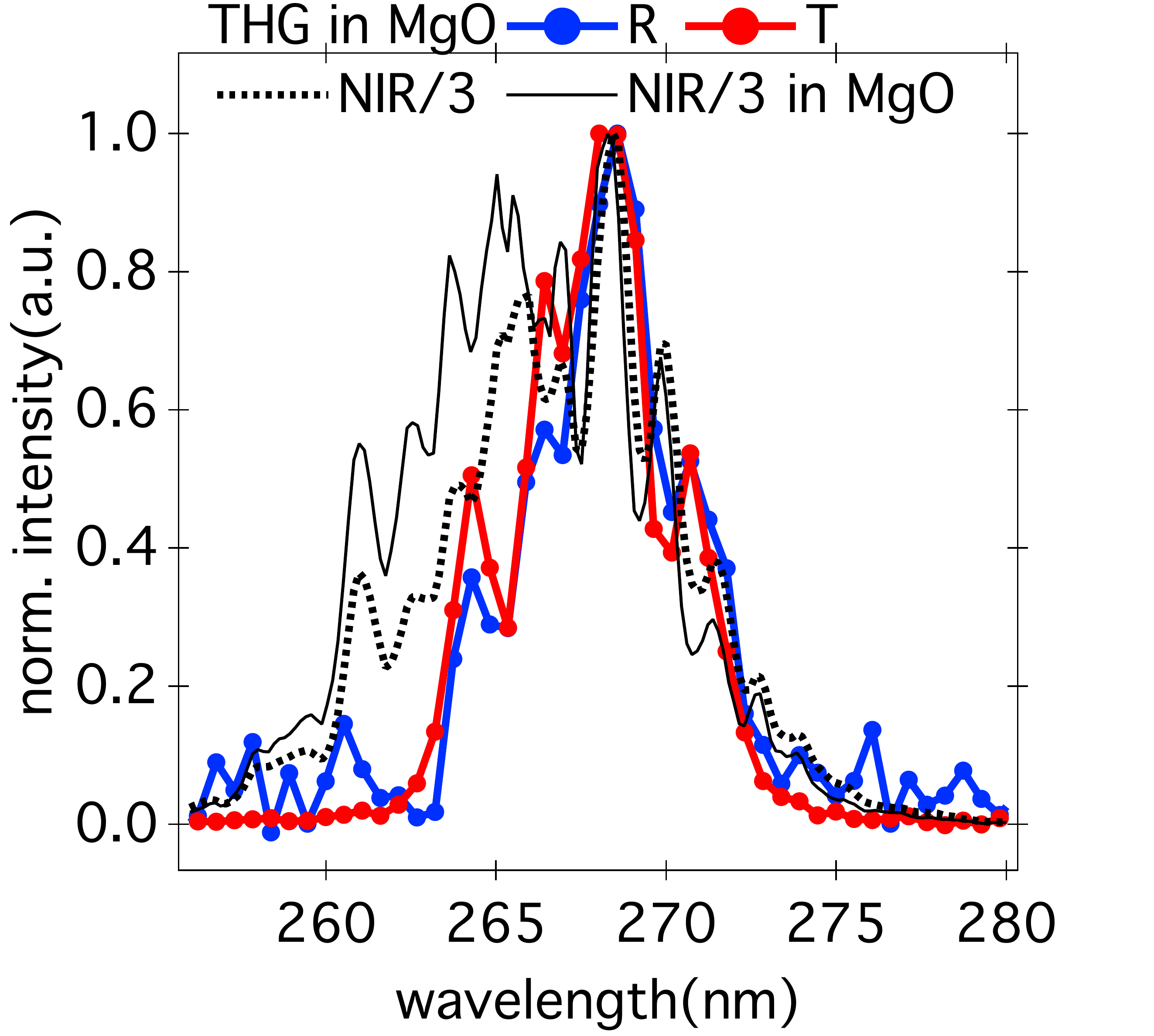}}}
	\subfigure[\label{fig:trans_CrMgO0}]{{\includegraphics[height=4.7cm,width=5cm]{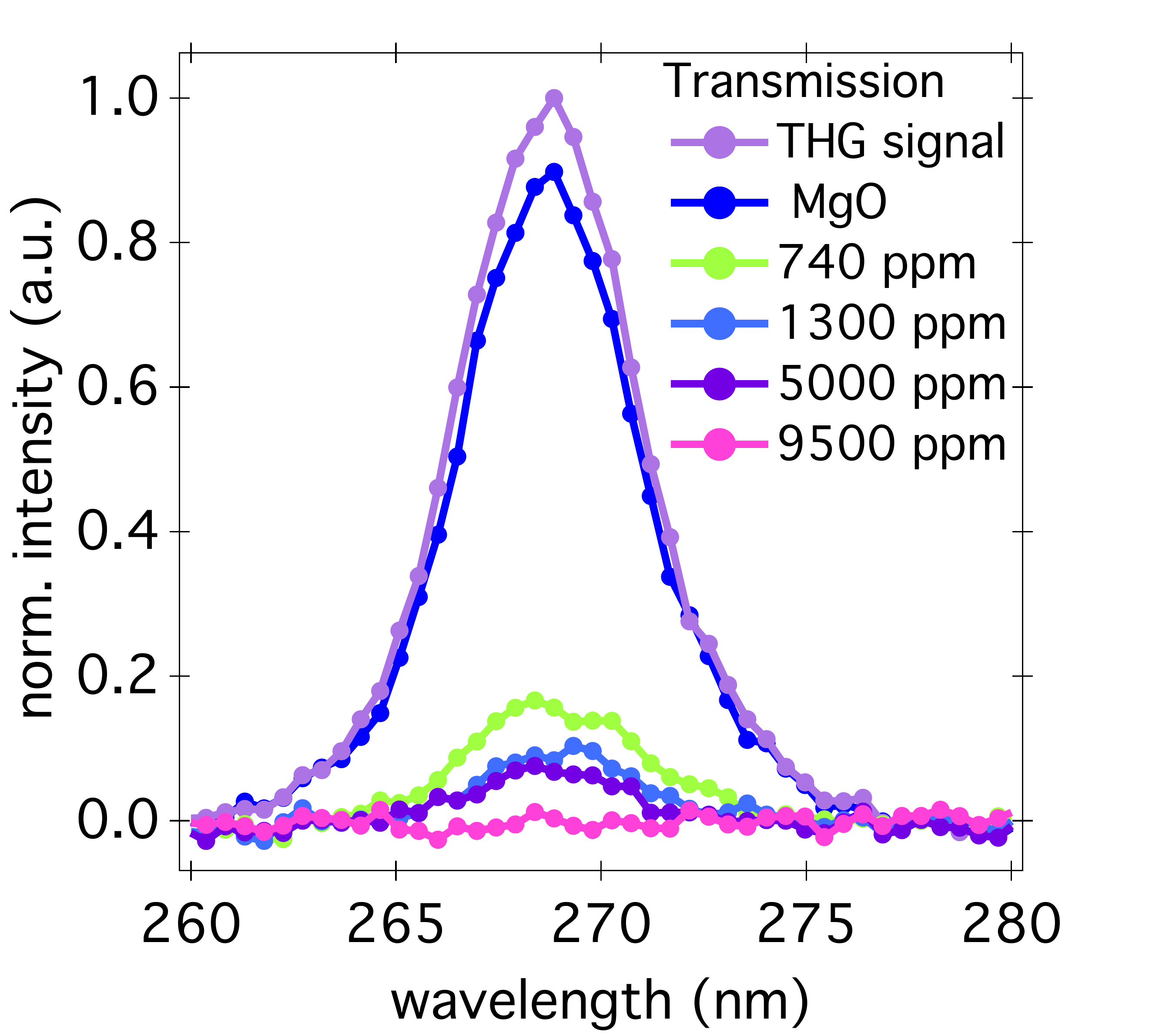}}}
	\caption{(a) Spectral measurements of THG in MgO (pure) and Cr: MgO with different doping concentrations at normal incidence of the driving field to crystals. At low and higher doping concentrations, the TH spectrum exhibits broadening and a redshift, (b) Spectral measurements of THG in transmission (blue dots) and reflection (red dots) in MgO. Where R: reflection and T corresponds to transmission. NIR/3 (black dotted in air and black line in MgO) to match the peak of TH  theoretically. (c) Transmission of THG signal through MgO and Cr: MgO crystals.}
\end{figure}

The spectral profile of THG in MgO and Cr: MgO crystals is shown in figure \ref{fig:THG_SpectrumT_CrMgO}. The peak of THG in MgO is centred at 266.5 nm with the FWHM of 4.25 nm. We have observed the spectral broadening of the TH signal is low (740 ppm) and higher doped (9500 ppm) crystal. The FWHM of TH from Cr: MgO (740 ppm) and Cr: MgO (9500 ppm) is broadened to 6.54 nm and 8.1 nm, respectively. The peak of TH in Cr: MgO (9500 ppm) is red-shifted centred at 268.5 nm. Whereas, the FWHM of TH signal generated in Cr: MgO (1300 ppm) and Cr: MgO (5000 ppm) reduces to 4 nm and 4.15 nm, respectively (figure \ref{fig:THG_SpectrumT_CrMgO}). The spectra of THG signal from MgO in transmission (T red dots) and reflection (R blue dots) measured with another UV-VIS spectrometer (Sarspec) is shown in figure \ref{fig:THG_trans_Ref_MgO}. The THG signal generated in transmission showed multiple pronounced peaks structure due to non-linear propagation effects than the TH signal observed from reflection geometry.

In the next step, we have generated a strong 267 nm signal in the SiO\textsubscript{2} crystal at an intensity of  $\sim$ 1.0 $\times$ $10^{13}$ W/cm\textsuperscript{2} to observe the transmission of 267 nm through these crystals. The transmitted spectral measurements of 267 nm through the MgO (pure) and different doped Cr: MgO crystals are shown in figure \ref{fig:trans_CrMgO0}. Minimal linear absorption of 267 nm by the MgO (pure) crystal is observed, while the transmission of 267 nm through Cr: MgO crystals decreases with the increase of doping concentration. There is a strong linear absorption of 267 nm by the Cr: MgO crystals. This shows that the THG from the front surface of the doped crystal will be absorbed and the observed third harmonic (TH) signal is obtained from the back surfaces in the doped crystal (Cr: MgO). The spectra of TH in low and higher doped crystal broaden due to the self-focusing effect which induces self-phase modulation of the driving pulses as observed in the figure  \ref{fig:IR_trans_Cr_MgO}.

\subsection{Propagation effects and generation of below band-gap harmonics in Sa}
\label{SHG_THG:Sa}

\begin{figure}[htbp]
	\subfigure[\label{fig:IR_trans_Ref_Sa}]{{\includegraphics[scale=0.19]{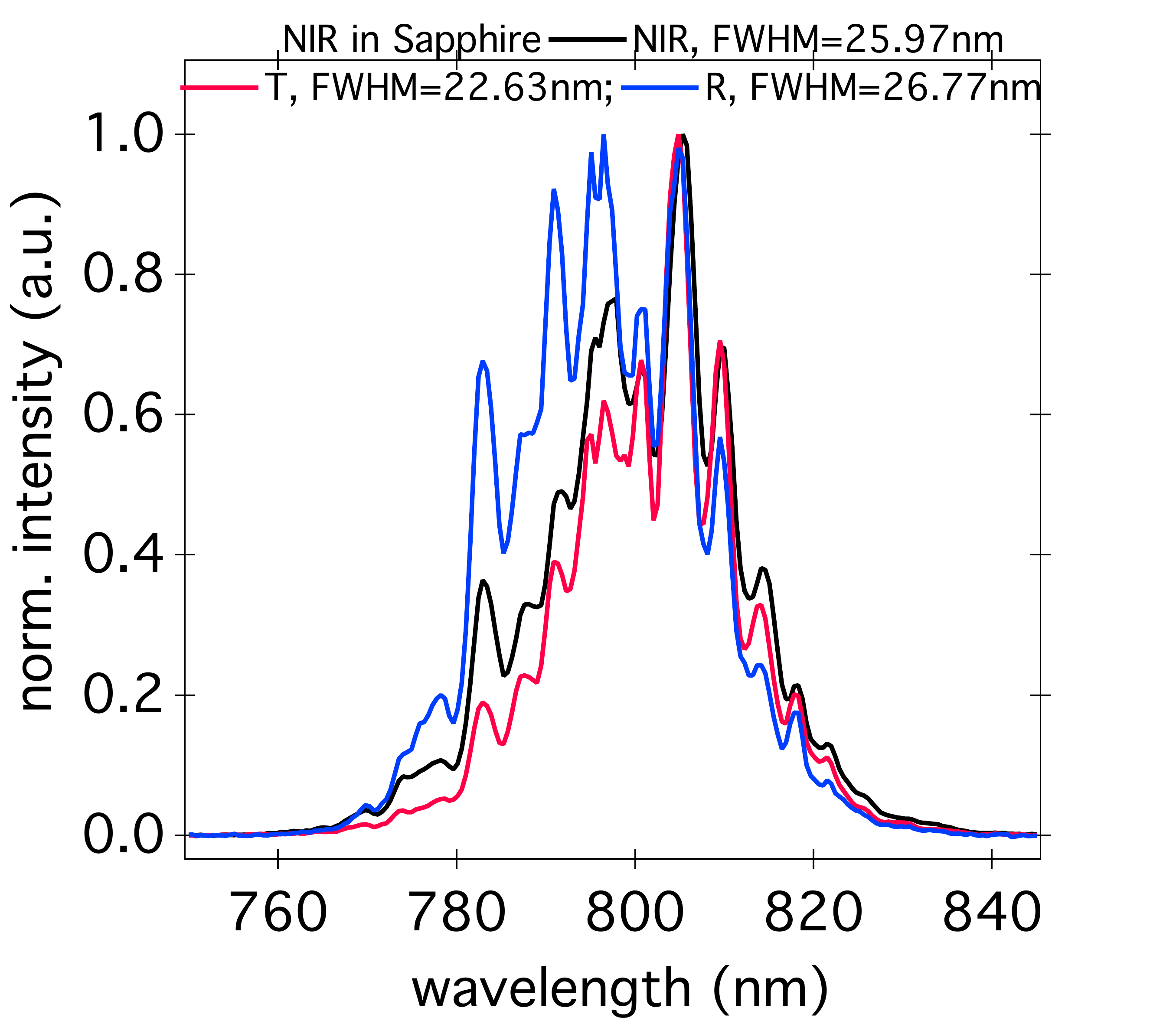}}}
	\subfigure[\label{fig:SHG_T_R_Sa}]{{\includegraphics[scale=0.19]{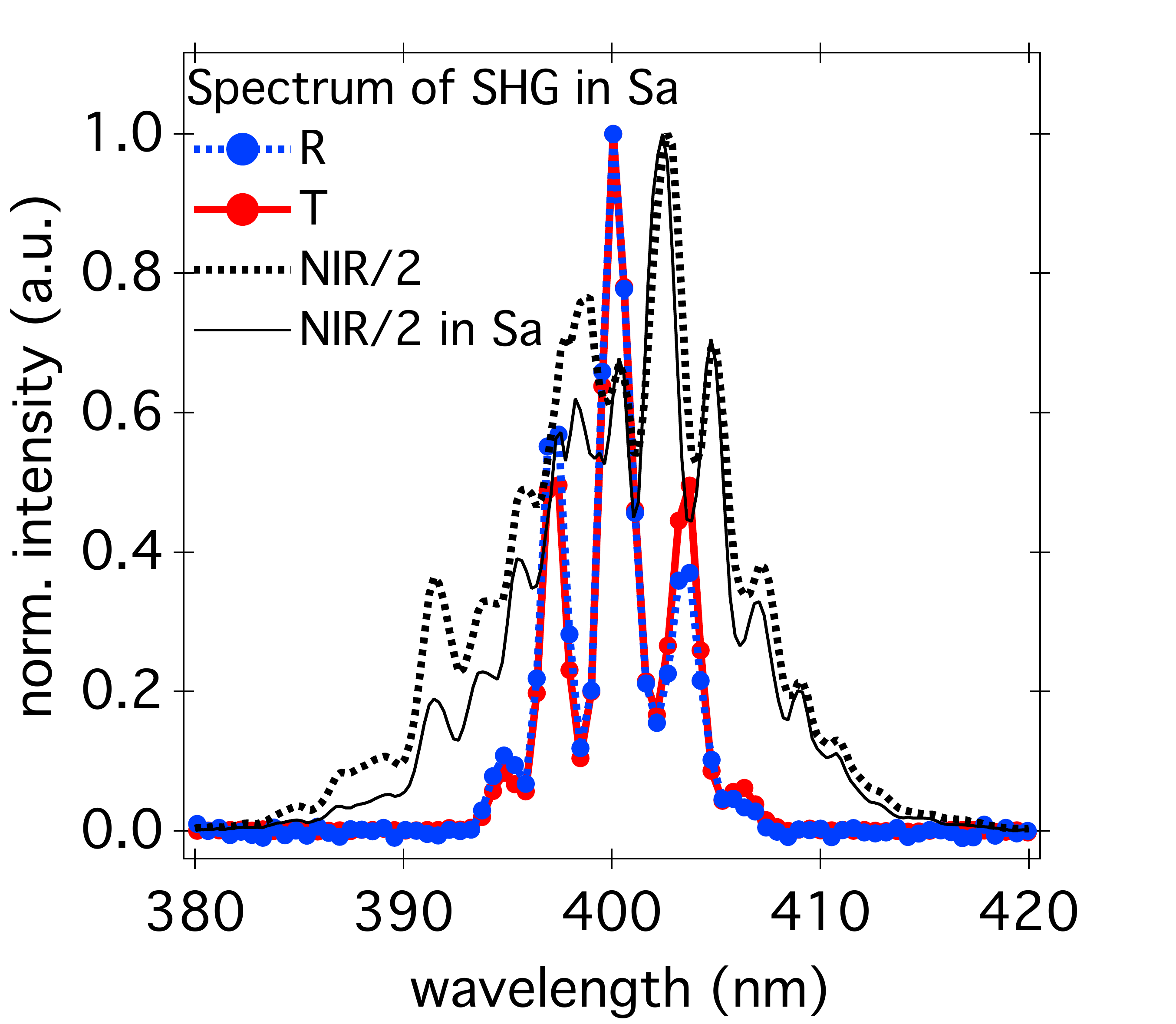}}}
	\subfigure[\label{fig:THG_T_R_Sa}]{{\includegraphics[scale=0.19]{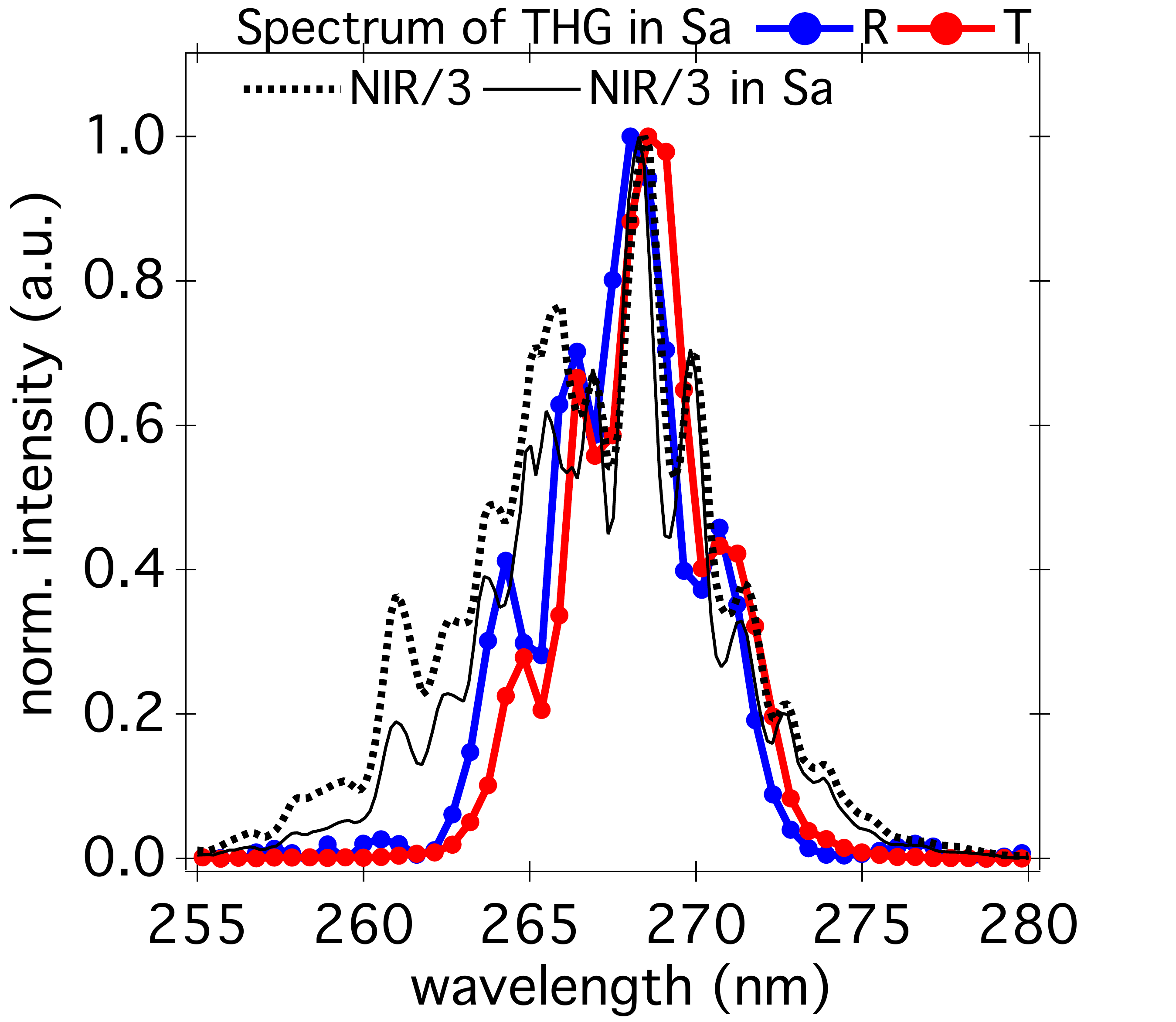}}}
	\caption{(a) Propagation of NIR (800 nm) in 300 $\mu$m  Sapphire (Sa) crystal. NIR ( black) FWHM=25.97 nm, transmission (red) and reflected from the front surface of Sa (Blue ) having FWHM=22.63 nm and FWHM=26.67 nm, respectively. (b) Spectral measurements of SHG in transmission (red data points) and reflection (blue data points) geometry. NIR/2 (black dotted in air and black line in Sa)  to match the peak of SH theoretically. (c) Spectral measurements of THG in reflection (blue data points) and in transmission(red data points). NIR/3 (black dotted in air and black line in Sa) to match the peak of TH  theoretically.  Where R: reflection and T corresponds to transmission.}
\end{figure}

In another experiment, we have observed the propagation effects of driving pulses and generated below bandgap harmonics (SHG and THG) in a Sa crystal. The experimental setup to observe propagation effects of driving pulses and generation of below bandgap harmonics is described in section \ref{exp:setup_PST}. The transmitted spectra of the driving field through Sa is compressed and has more enhancement towards the leading edge of spectra and the trailing edge of spectra falls as compared to the driving NIR spectrum. As a result, the FWHM of NIR transmitted through Sa is 22.63 nm while the FWHM of NIR is 25.97 nm and reflected FWHM of NIR from Sa is 26.77 nm which is more blue-shifted and has pronounced peaks at the trailing edge figure \ref{fig:IR_trans_Ref_Sa}. 

We have generated SHG and THG in Sa (Al$_2$O$_3$), oriented along (0001) having dimension $10\times10\times0.3$ mm$^{3}$ at $\sim$1.0 $\times$ $10^{13}$ W/cm\textsuperscript{2} . The acquisition time for SHG 100 ms and for THG chosen to 1s with average over five acquisitions to measure the spectrum of THG by UV-VIS spectrometer. The spectral measurements of SHG in Sa in transmission and reflection geometry is shown in figure \ref{fig:SHG_T_R_Sa}. The SH spectra has the FWHM of 6.42 nm  in transmission whereas in reflection geometry has FWHM of 6.27 nm. The spectral measurements of THG in transmission and reflection geometry is shown in figure \ref{fig:THG_T_R_Sa}. FWHM of TH signal in transmission is 5.43 nm while in reflection geometry has the FWHM of 5.35 nm. The THG signal generated in transmission is broader  as compared to reflection in Sa (figure \ref{fig:THG_T_R_Sa}).

\begin{figure}[htbp]
	\subfigure[\label{fig:Sa_SHG_Polar}]{{\includegraphics[scale=0.25]{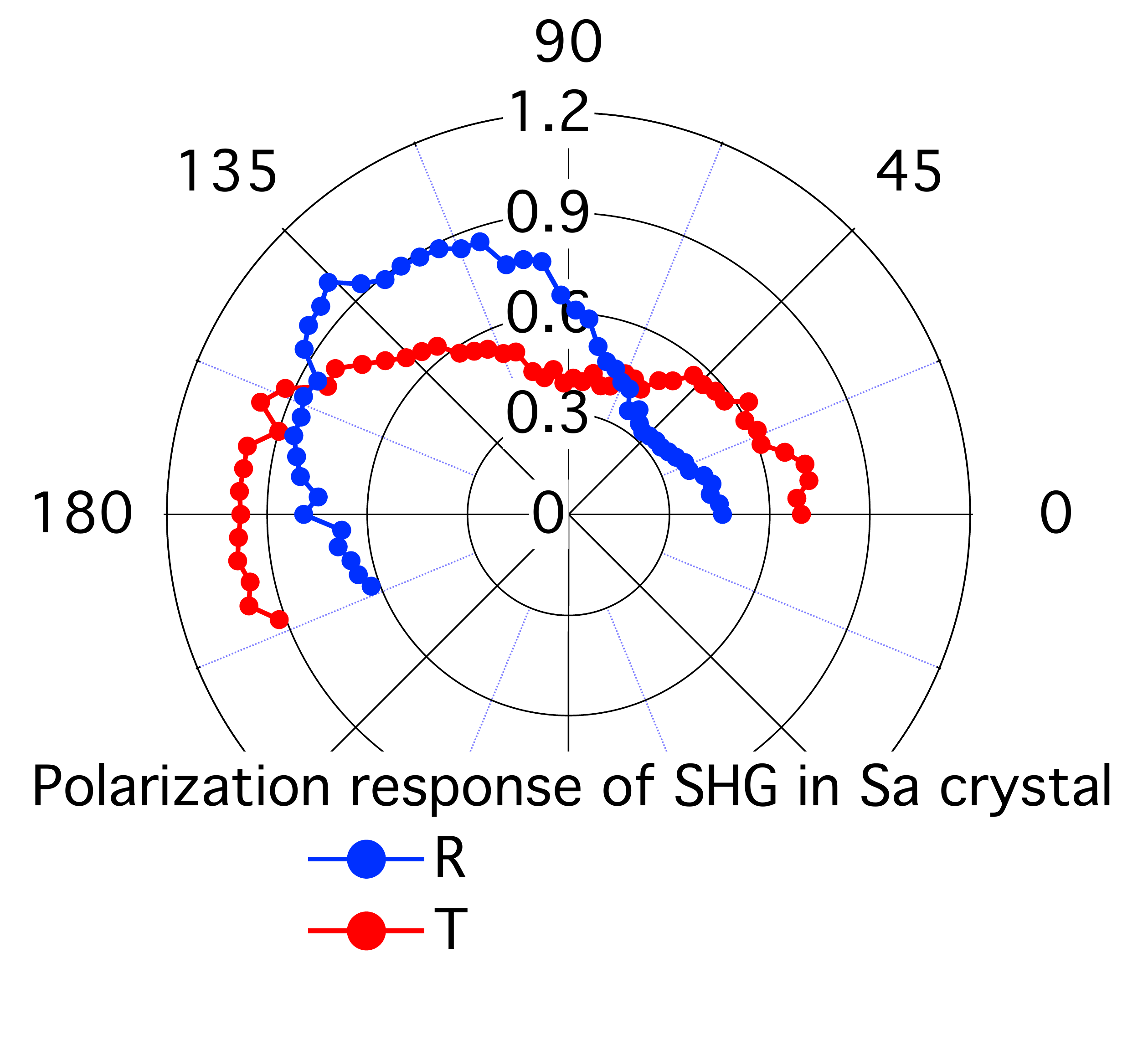}}}
	\subfigure[\label{fig:Sa_THG_Polar}]{{\includegraphics[scale=0.25]{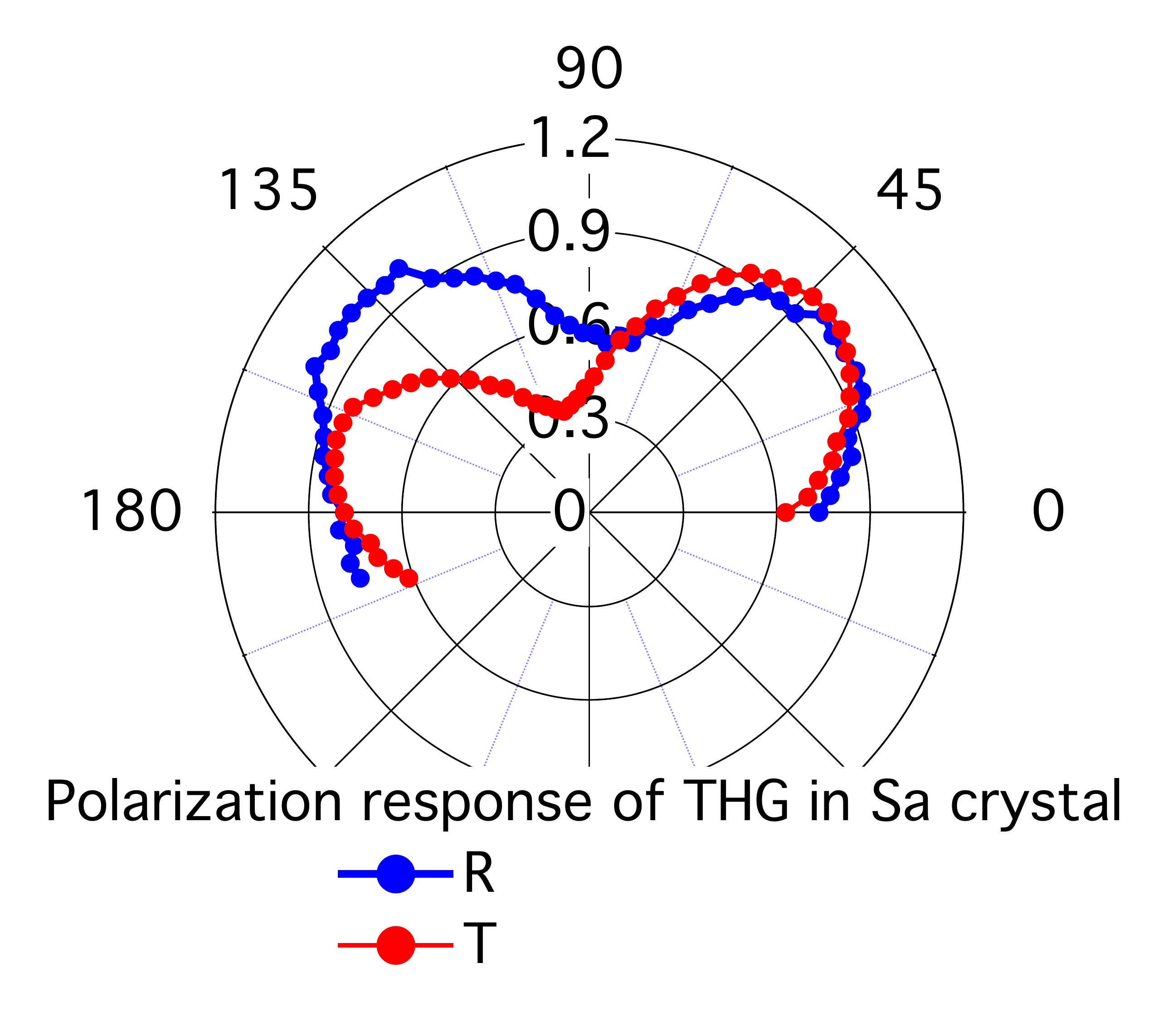}}}
	\caption{ Polarization dependence of harmonics in Sa in reflection and transmission geometry. Where, R: reflection and T corresponds to transmission (a) SHG, (b) THG.}
\end{figure}

The polarization response of SHG in transmission (shown in red dots) and reflection (blue dots)  in Sa demonstrated the two-fold symmetry with more isotropic response while an anisotropic response in reflection (blue dots) as shown in figure \ref{fig:Sa_SHG_Polar}. Whereas, a four-fold anisotropic polarization response of THG in Sa in transmission (red dots) and reflection geometry (blue dots) is observed as shown in figure \ref{fig:Sa_THG_Polar}. The polarization response of THG in both reflection and transmission agreed with different crystal cuts as reported earlier \cite{yi2017nonlinear}. The peaks of polarization dependence of THG in reflection are symmetric in intensity whereas an asymmetric response of peaks observed in the transmission is attributed to the non-linear propagation effects of the driving pulses in Sa.

\subsection{Propagation effects and generation of below band-gap harmonics in FS}
\label{SHG_THG:FS}

\begin{figure}[htbp]
	\centering
	\subfigure[\label{fig:IR_trans_Ref_FS}]{{\includegraphics[height=4.6cm, width=4.6cm]{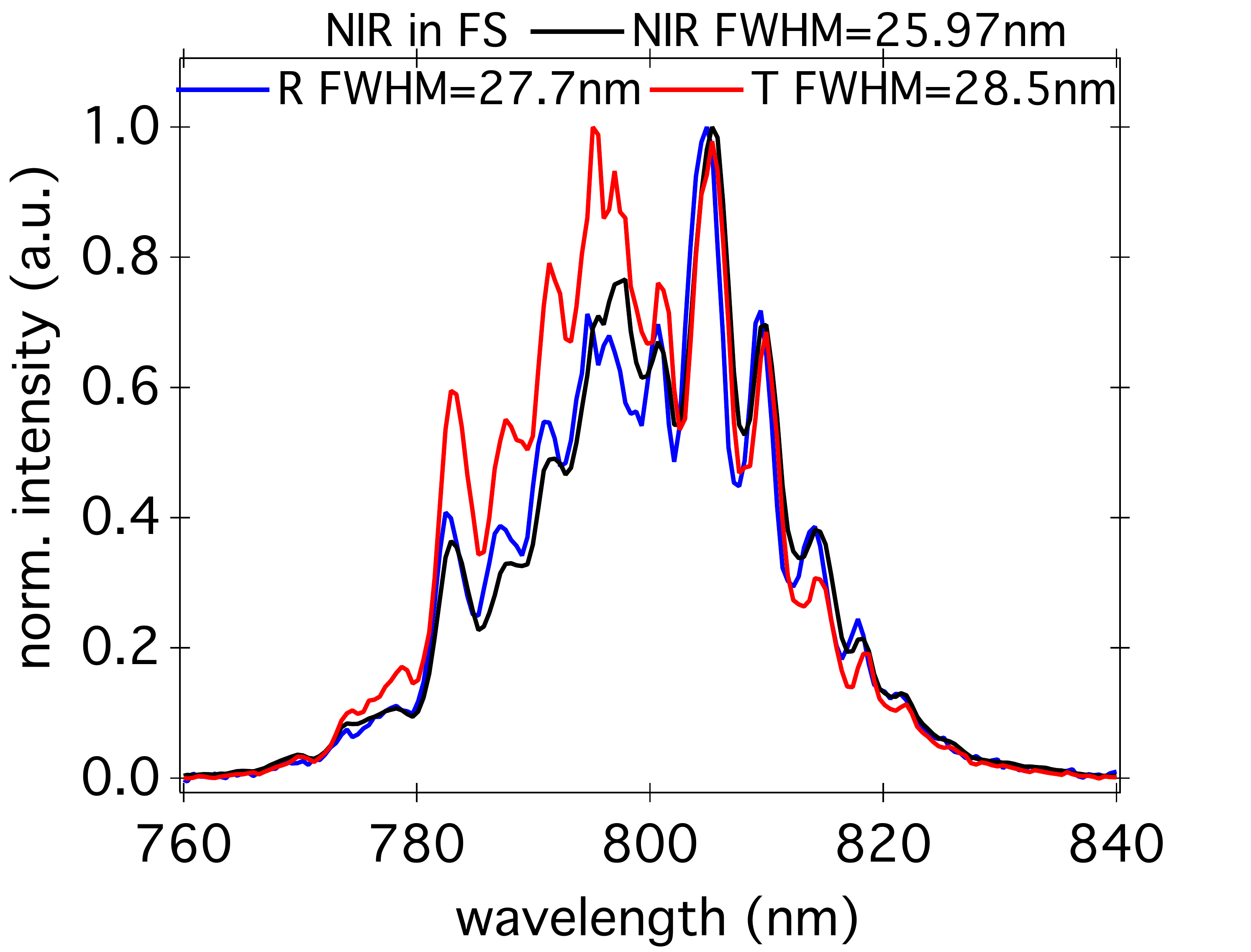}}}
	\subfigure[\label{fig:SHG_FS}]{{\includegraphics[scale=0.2]{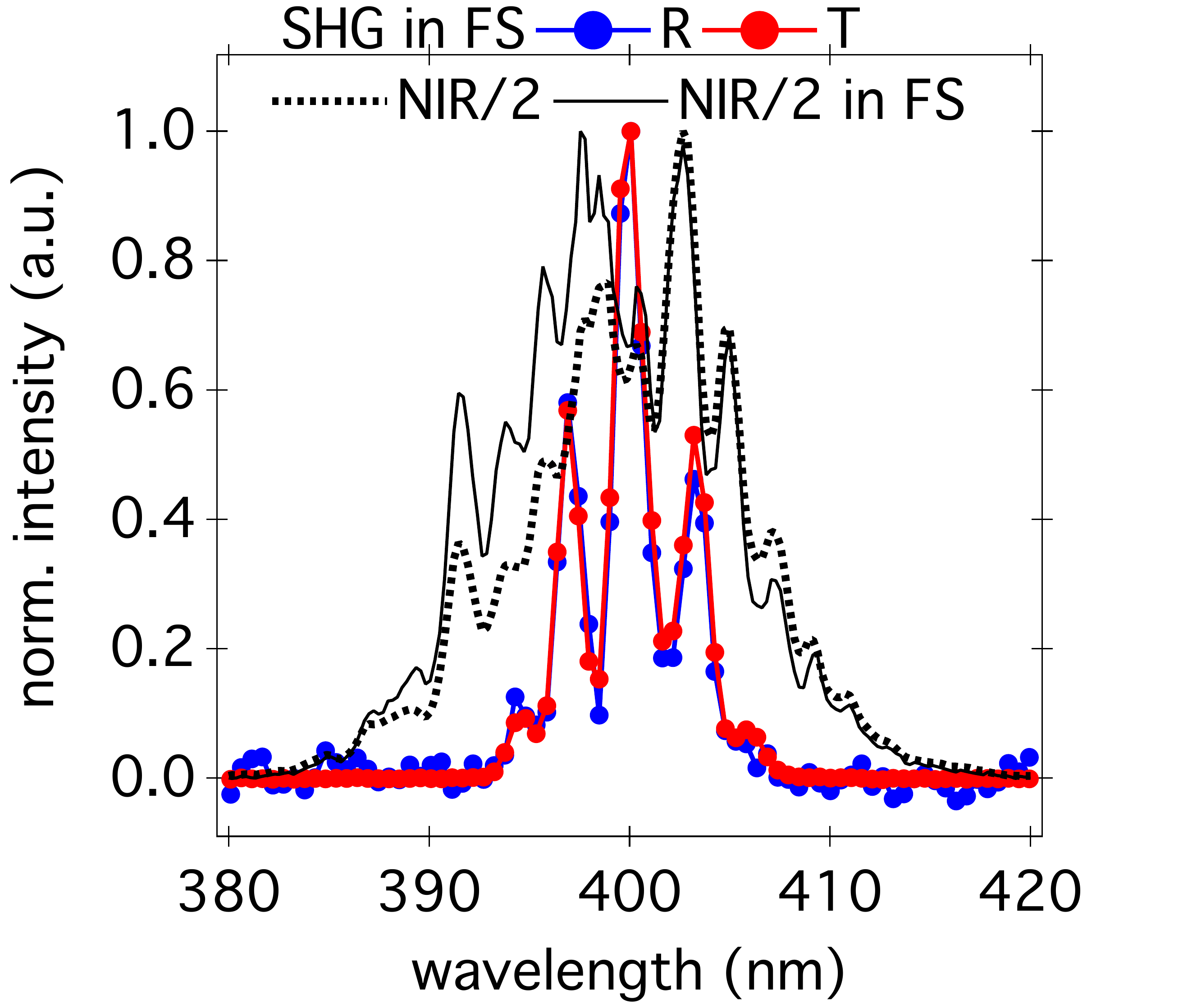}}}
	\subfigure[\label{fig:THG_FS}]{{\includegraphics[scale=0.2]{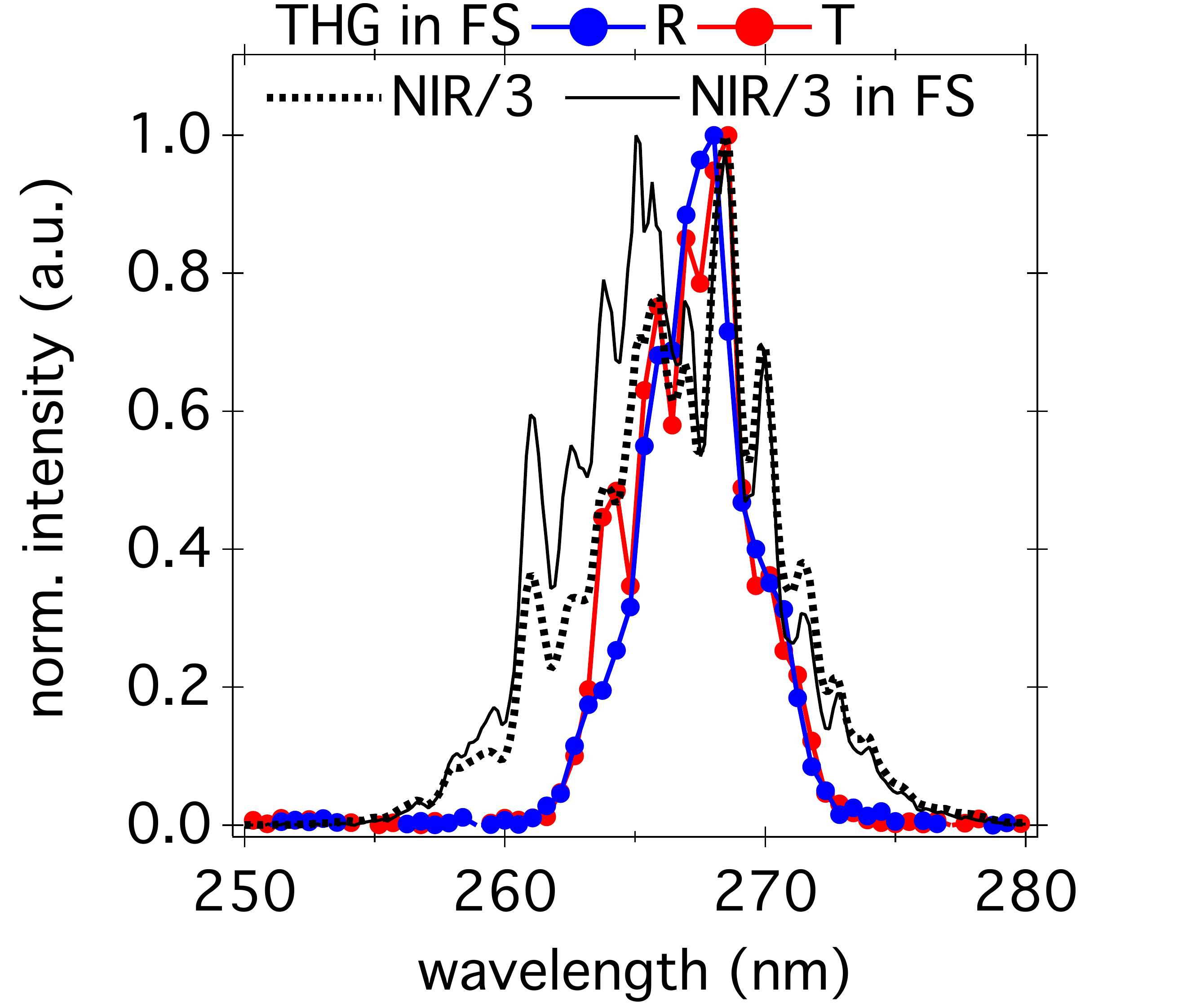}}}
	\caption{(a) NIR (black) after transmission through FS (red) and reflection (blue) from FS. The FWHM of transmitted NIR is 28.5 nm and reflected NIR is 27.7 nm, respectively. Whereas the FWHM of fundamental NIR is 25.97 nm. (b) Spectral measurements of SHG in 300 $\mu$m thick fused silica (FS) in transmission (red data points) and reflection (blue data points) geometry. NIR/2 (black dotted in air and black line in FS) to match the peak of SH theoretically. (c) Spectral measurements of THG in FS in reflection (blue data points)  and transmission(red data points). NIR/3 (black dotted in air and black line in FS) to match the peak of TH theoretically.}
	\label{fig:FS}
\end{figure}

The FS mounted on the translation stage to translate at the focus point of the driving field to observe the propagation effects of driving pulses. The broadening of driving pulses observed in FS, FWHM of NIR driving field in transmission (28.5 nm) and reflection (27.7 nm) is observed as compared to the fundamental FWHM (25.97 nm) as shown in figure \ref{fig:IR_trans_Ref_FS}. Due to non-linear propagation effects, the transmitted spectra (red) showed multiple enhanced peak structures at the trailing edge as compared to the fundamental driving field  (black curves). We have observed the blue-shifted broadening of the driving pulses is transmitted or reflected from the FS which attributes to the strong photoionization effect. Furthermore, SH and TH have been generated in 300 $\mu$m thick FS (detail of experimental setup is described in section \ref{exp:setup_PST}). The spectral profile of SHG  in transmission (red data points) and reflection (blue data points) is shown in figure \ref{fig:SHG_FS}. We have observed fringes in the SHG spectrum both in reflection and transmission which are separated by 3.15 nm. These fringes are imprinted from the driving pulses as shown in black dotted (NIR/2) in figure \ref{fig:SHG_FS}. Noted that the central peak of the SHG spectrum is blue-shifted when compared with the theoretically driving SH spectrum (NIR/2 black dotted spectrum). The broadening effect has observed in the TH in transmission as compared to the reflected TH signal in FS. We have observed that the spectral profile of TH in reflection is blue-shifted than in transmission.

\begin{figure}[htbp]
	\subfigure[\label{fig:SHG_FS_Polar}]{{\includegraphics[scale=0.25]{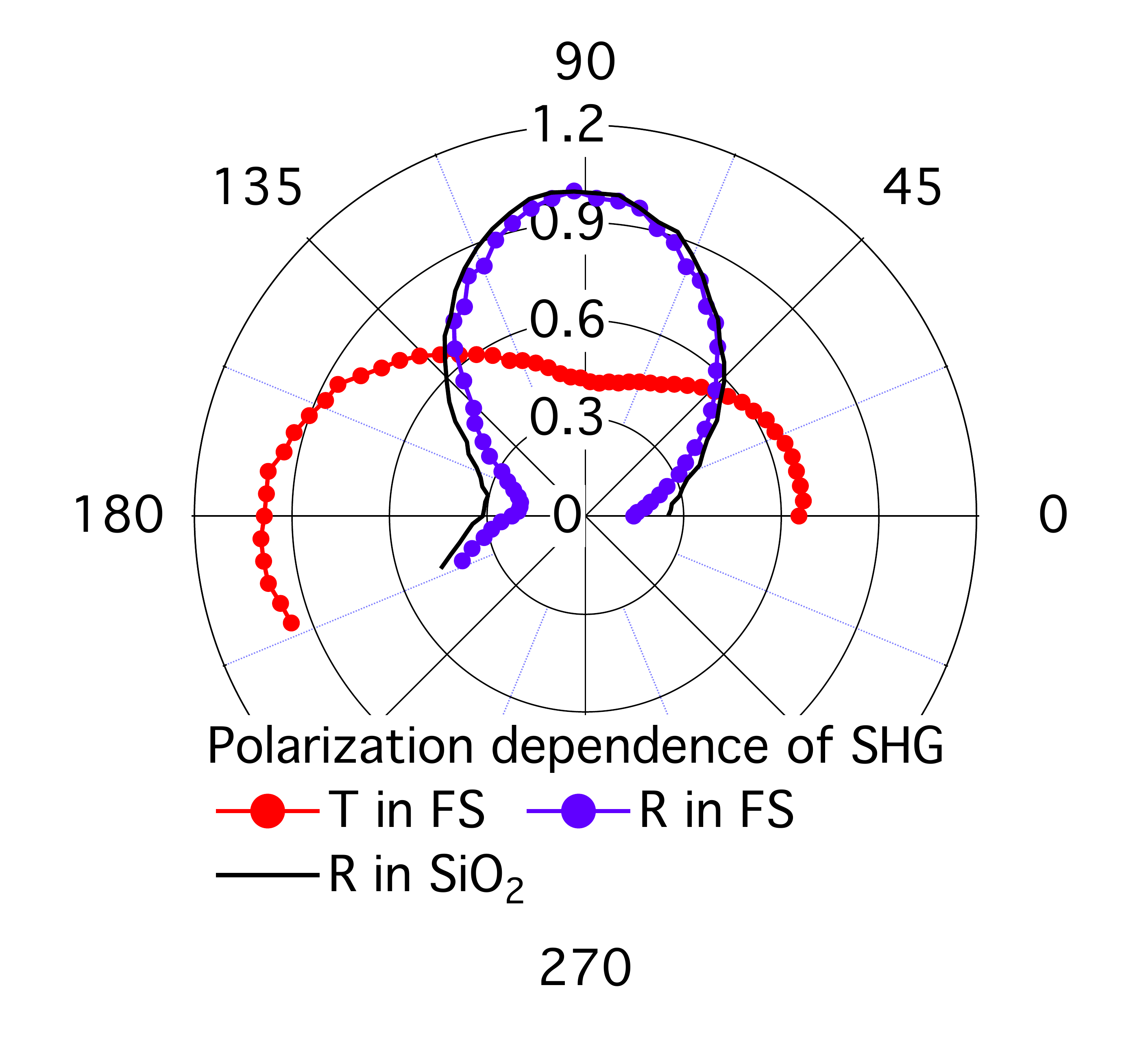}}}
	\subfigure[\label{fig:THG_FS_Polar}]{{\includegraphics[scale=0.22]{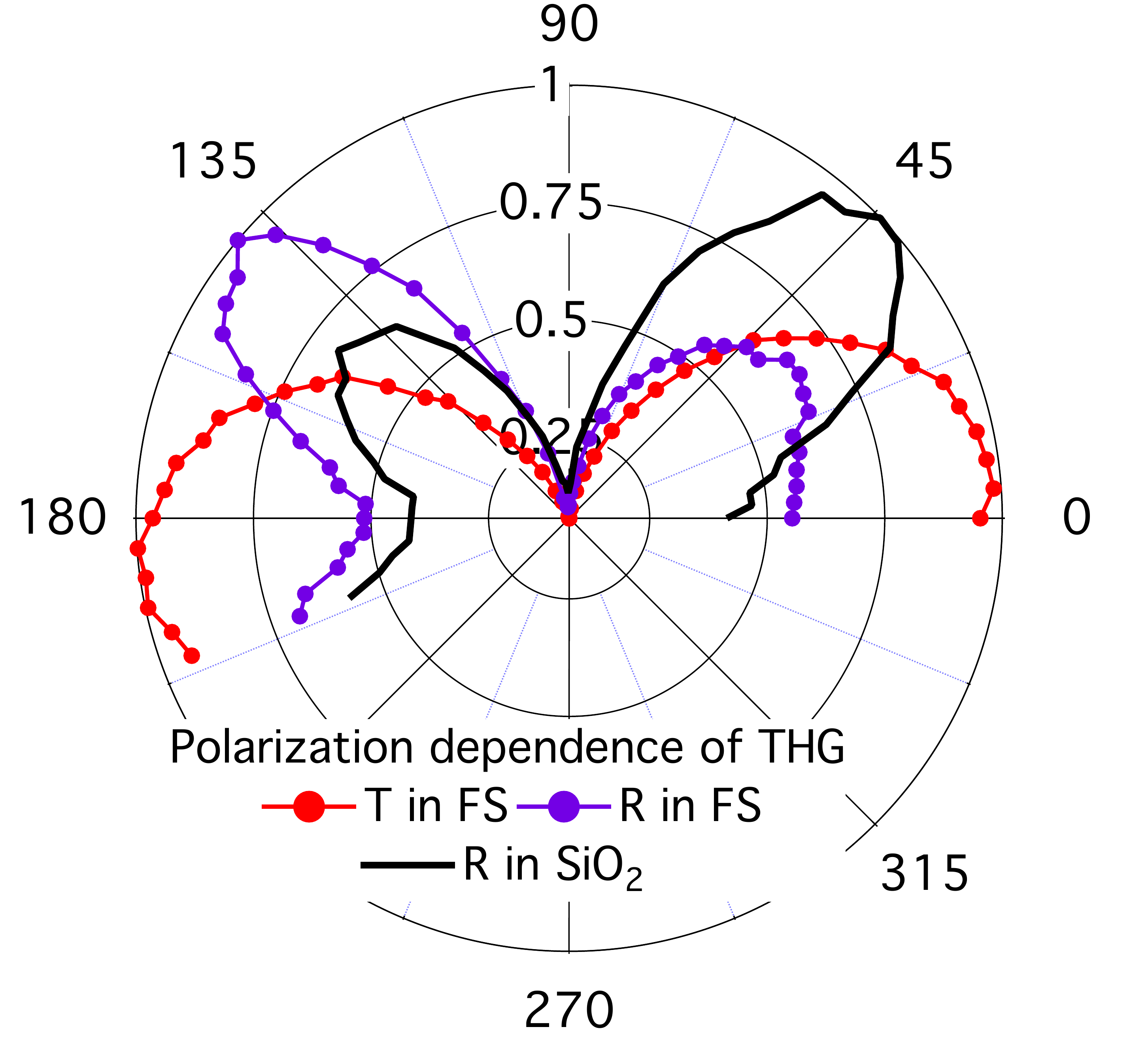}}}
	\caption{ Polarization dependence of harmonics in FS in reflection and transmission geometry (a) SHG, (b) THG.}
\end{figure}

The polarization of driving pulses is rotated by HWP to observe the polarization dependence of SH and TH in both reflection and transmission geometry.  We have observed the isotropic response of SHG in FS in transmission geometry (red data points) and two-fold anisotropic response in reflection as shown in figure \ref{fig:SHG_FS_Polar}. We have observed that the symmetry response of SH in reflection and transmission are orthogonal.  In reflection geometry, SHG shows more anisotropic response and has a two-fold symmetry along the vertical linear polarization direction. This anisotropic response of SHG in reflection geometry of FS is similar to crystalline SiO$_2$ response as shown in figure \ref{fig:SHG_FS_Polar}.
Although, FS is amorphous (non-crystalline) in nature and does not exhibit any long-range symmetry. Yet, the THG shows two fold-symmetry as illustrated in figure \ref{fig:THG_FS_Polar} (red data points) in transmission. The intensity of THG is maximum for the linear horizontal polarization and minimum generation efficiency along with the vertical linear polarization. Surprisingly, the polarization response of THG (dark purple data points) in reflection geometry in  FS demonstrate a highly anisotropic four-fold response which has similar to the crystalline SiO$_2$ (figure \ref{fig:THG_FS_Polar}). Although the amorphous quartz (FS) lacks the long-range periodicity but still can exhibit strong polarization dependence of below bandgap harmonics in reflection (four-fold)) and transmission (two-fold). We attributed this to the localization of electron excursion distance in amorphous and crystalline SiO$_2$ \cite{you2017high}. The polarization response of SHG and THG of FS  in reflection is similar to the crystalline SiO$_2$ (black solid line) (figure \ref{fig:THG_FS_Polar}) which we attribute to the crystalline nature of FS at its surface.

\section{Conclusion}
\label{conclusion}

We have investigated the non-linear propagation effects of the ultrafast intense near-infrared driving field at 800 nm of 40 fs duration operating at a repetition rate of 1 kHz through the wide bandgap dielectrics. We observe a blue-shifted broadening in the driving pulse in pristine MgO and overall broadening observed in Cr doped MgO crystals. Below bandgap harmonics (SHG and THG) in thin solids has generated to explore the non-linear response at strong fields.  The non-linear propagation effects are avoided in low-order harmonics through reflection geometry and compared with the transmission harmonics. We attribute the broadening of driving pulses in transmission due to the self-phase modulation. The observed propagation effects are imprinted on the below bandgap harmonics. The polarization response of harmonics in Sa and FS are more anisotropic in reflection geometry than generated through the transmission. This work shows the sensitivity to control the spectral profile of harmonics by manipulating the driving field, showing the possibility of new tailored solid-state XUV sources for optical diagnostics.

\section*{Acknowledgements}
This work was supported by the Foundation for Science and Technology (FCT) under the grant number PD/BD/135224/2017 in the framework of the Advanced Program in Plasma Science and Engineering (APPLAuSE).

\clearpage

\end{document}